\documentclass[aps,pra,twocolumn,tighten,groupedaddress,floatfix]{revtex4-1}

\usepackage{graphicx}
\usepackage{dcolumn}
\usepackage{bm}
\usepackage{amsmath}
\usepackage{xcolor}
\usepackage{braket}

\usepackage{amsfonts}
\usepackage{blkarray}
\usepackage{amssymb}
\usepackage{multirow}
\usepackage{mathrsfs}

\newcommand{\fdlu}{$\text{1}\!\downarrow\!\text{2}\!\uparrow\,$}
\newcommand{\fuld}{$\text{1}\!\uparrow\!\text{2}\!\downarrow\,$}
\newcommand{\fulu}{$\text{1}\!\uparrow\!\text{2}\!\uparrow\,$}

\begin{document}

\title{Third-order momentum correlation interferometry maps for entangled quantal states of three singly 
trapped massive ultracold fermions}

\author{Constantine Yannouleas}
\email{Constantine.Yannouleas@physics.gatech.edu}
\author{Uzi Landman}
\email{Uzi.Landman@physics.gatech.edu}

\affiliation{School of Physics, Georgia Institute of Technology,
             Atlanta, Georgia 30332-0430}

\date{23 February 2019}
\begin{abstract}
Analytic higher-order momentum correlation functions associated with the time-of-flight spectroscopy of three 
ultracold fermionic atoms singly-confined in a linear three-well optical trap are presented, corresponding to 
the $W$- and Greenberger-Horne-Zeilinger-type $(GHZ)$ states that belong to characteristic classes of 
tripartite entanglement and represent the strong-interaction regime captured by a three-site Heisenberg 
Hamiltonian. The methodology introduced here contrasts with and goes beyond that based on the standard Wick's 
factorization scheme; it enables determination of both third-order and second-order spin-resolved and 
spin-unresolved momentum correlations, aiming at matter-wave interference investigations with trapped massive
particles in analogy with, and having the potential for expanding the scope of, recent three-photon 
quantum-optics interferometry.
\end{abstract}

\maketitle

\section{Introduction}
\label{int}
 
Matter-wave simulations, with highly-controlled ultracold atoms, of well-known photon 
physics have been pursued along two quantum-optics central themes: (i) the coherence properties 
\cite{glau63,glau06} of thermal or chaotic light (in contrast to laser light), studied via second- and 
higher-order correlations (including the Hanbury Brown-Twiss effect \cite{twis56}), and (ii) two-photon (or 
biphoton) interference effects \cite{mand99,shihbook,oubook} associated with fully quantal and entangled photon
states (including the Hong-Ou-Mandel effect \cite{hong87}).

Knowledge of high-order correlations of a quantum many-body system has been long recognized to fully 
characterize the system under study \cite{schw51.1,schw51.2,schm17,kher17,glau63}. Most recently progress has 
been demonstrated \cite{bran17,bran18,yann18,yann19} in the development of matter-wave interferometry 
through the use of second-order momentum correlations, 
\textcolor{black}{
measureable in time-of-flight (TOF) laboratory experiments \cite{note2,note1,berg19,prei19}, 
}
yielding exact closed-form results based on first-principles 
(configuration interaction \cite{bran17,bran18}) and model Hamiltonian (Hubbard \cite{yann18,yann19}) methods.

Here we formulate and implement an accurate and practical methodology for determining higher-order momentum 
correlation functions for strongly interacting and entangled many-particle systems (beyond the bosonic or 
fermionic quantum-statistics entanglement contributions), expanding and generalizing the above-mentioned work
\cite{bran17,bran18,yann18,yann19}. 
\textcolor{black}{
In particular, our present methodology and derivation of higher-order 
momentum correlations (here, spin-resolved and spin-unresolved third-order correlations) based on the 
Heisenberg Hamiltonian for three singly-trapped ultracold atoms, differs from that relying on the standard 
Wick's factorization scheme \cite{wick50}. 
That latter scheme is central to investigations of many-particle 
correlations in varied fields (including nuclei, condensed matter, atoms and molecules \cite{note1,prei19}
and optics), allowing, {\it in the absence of interactions}, full factorization (with the use of the 
Wick method \cite{wick50,zinnbook} of the $N$-particle correlation function (Green's function in the original 
formulation \cite{wick50}), ${\cal G}^N$, $N>2$, as a sum of terms containing antisymmetrized/symmetrized 
(corresponding to fermions/bosons) products of only ${\cal G}^N$'s with $N \leq 2$.}   

\textcolor{black}{
The Wick's factorization has been employed for Gaussian-type, or single-determinantal, ground states of ultracold 
atomic clouds \cite{hodg11,hodg11b,hodg13,west06,prei19,aspe19}, 
}
mimicking the methodology, introduced earlier 
\cite{glau63,glau06} for addressing coherence properties of thermal or chaotic light, which was not focused
on quantal effects (such as entanglement) at zero temperature. In contrast, these fundamental quantum effects, 
which are targeted (see, e.g., \cite{isla15}) in current ultracold atom research relating to fundamentals of 
quantum information are central to our present work.

\textcolor{black}{
Indeed, in light of the limitation of the standard Wick's method \cite{cede74,wagn91,stef12} to determinantal 
spin-non-degenerate ground states (being restricted to the highest-spin fermionic component \cite{cede74,prei19}
or to spinless bosons \cite{aspe19}), and thus the inability of that scheme to treat spin-degenerate ground-states  
(ubiquitous in investigations of quantum chemistry, condensed-matter, and quantum information, e.g.,  the $W$ and 
$GHZ$ states studied herein), our methodology and the results we uncovered (including the highlighting and 
demonstration of the important role of spin-resolved momentum correlations), open avenues for analysis, 
characterization, and understanding of recent and ongoing experiments (particularly TOF of trapped, interacting, 
ultracold atoms) with a focus on relevant highly-entangled states as a resource in quantum information.  
}

To put this development in context, we note here recent progress in the experimental processing of data and 
control and manipulation of ultracold atoms in colliding free-space beams or clouds (including free fall under 
the cloud's gravity) \cite{aspe07,hodg11,hodg13,kher13,aspe15,kher17,schm17} or in optical traps and tweezers 
({\it in situ\/} or TOF) \cite{foel05,kauf14,joch15,kauf18}, which has motivated a growing number of both 
experimental \cite{aspe07,hodg11,kher13,aspe15,kher17,schm17,foel05,kauf14,joch15,kauf18} and theoretical 
\cite{kher14,bran17,bran18,bonn18,yann18,yann19} studies concerning the analogies between quantum optics and 
matter-wave spectroscopy.

The paper is organized as follows: In Section II, we outline the three-site Heisenberg model and its solutions. 
Section III presents background material for the many-body methodology used for obtaining the momentum correlation 
functions, whereas Section IV gives results for the $W$ states. The cases of spin {\it un\/}resolved momentum 
correlations for the $W$ states are presented in Sect. IV.A (third order) and in Sect. IV.B (second order). 
Spin resolved momentum correlations for the $W$ states are discussed in Sect. IV.C. (third order), 
and in Sect. IV.D. (second order). Results for the momentum correlation functions for the $GHZ$ state are 
discussed in Section V. Our conclusions are given in Section VI. 

\section{Outline of three-site Heisenberg model and its solutions} 

The three-fermion $|W\rangle$ and $|GHZ\rangle$ strongly
entangled three-qubit states \cite{zeil90,cira00} that are the focus of this paper are solutions \cite{raja02}
of the following three-site linear-spin-chain Heisenberg Hamiltonian (which describes the strong-interaction
limit of the Hubbard model \cite{auerbook}) 
\begin{align}
H=(J/2) ({\bf S}_1 \cdot {\bf S}_2 + {\bf S}_2 \cdot {\bf S}_3) - J/2,
\label{hh} 
\end{align} 
where $J$ is the exchange coupling between sites and ${\bf S}_i$ is the spin operator of the particle 
associated with the $i$th site. 

First we will address the case of the $W$ states, which are the $S_z=1/2$ eigenstates of the above 
Heisenberg Hamiltonian $H$ \cite{raja02}. 

Using the three-member ket-basis $|\uparrow \uparrow \downarrow\rangle$,
$|\uparrow \downarrow \uparrow \rangle$, and $|\downarrow \uparrow \uparrow \rangle$, the above Hamiltonian
is written in matrix form
\begin{align}
H = \frac{J}{2} \left( 
\begin{array}{ccc}
-1 & ~1 & ~0 \\
~1 & -2 & ~1 \\
~0 & ~1 & -1 \\
\end{array} \right).
\label{hm}
\end{align}

The eigenvalues of the matrix in Eq.\ (\ref{hm}) are:
\begin{align}
\begin{split}
&{\cal E}_1 = -3J/2,\;\; S=1/2,\\
&{\cal E}_2 = -J/2,\;\; S=1/2, \\
&{\cal E}_3 = 0,\;\; S=3/2.
\label{ee}
\end{split}
\end{align}

The corresponding (normalized) eigenvectors and their total spins are given by:
\begin{align}
\begin{split}
&{\cal V}_1=
\{1/\sqrt{6}, -1/\sqrt{3}, 1/\sqrt{6} \}^T,\;\;S=1/2,\\
&{\cal V}_2=
\{-1/\sqrt{2}, 0, 1/\sqrt{2} \}^T,\;\;S=1/2,\\
&{\cal V}_3=
\{1/\sqrt{3}, 1/\sqrt{3}, 1/\sqrt{3} \}^T,\;\;S=3/2.
\label{vv}
\end{split}
\end{align}

\section{Many-body methodology for momentum correlations: Preliminaries}

To generate the third-order momentum correlation maps ${\cal G}_i^3(k_1,k_2,k_3)$, $i=1,2,3$, corresponding to 
the three $W$-type solutions in Eq.\ (\ref{vv}) of the Heisenberg Hamiltonian, one needs to transit to the 
first-quantization formalism using  momentum-dependent Wannier-type spin-orbitals.
To this effect, each fermionic particle in any of the three wells is represented by a displaced Gaussian 
function \cite{bran17,bran18,yann19}, which in momentum space is given by
\begin{equation}
\psi_{j}(k)\chi(\omega) = \frac{2^{1/4}\sqrt{s}}{\pi^{1/4}} e^{-k^2 s^2} e^{i d_j k} \chi(\omega).
\label{psikd}
\end{equation}
In Eq.\ (\ref{psikd}), $d_j$ ($j=1,2,3$) denotes the position of each of the three wells, $s$ is the width of 
the Gaussian function. 

$\chi(\omega)$ is a shorthand notation for the spin-up, $\alpha(\omega)$, or spin-down, $\beta(\omega)$, 
single-particle spin functions. The two spin functions are orthonormal according to the formal way 
\cite{szabobook} $\int d\omega \alpha^*(\omega) \alpha(\omega) = \int d\omega \beta^*(\omega) \beta(\omega)
=1$, $\int d\omega \alpha^*(\omega) \beta(\omega) = \int d\omega \beta^*(\omega) \alpha(\omega) =0$. 

Employing the fact that in the first-quantization representation the basis kets, 
$|\uparrow \uparrow \downarrow\rangle$,
$|\uparrow \downarrow \uparrow \rangle$, and $|\downarrow \uparrow \uparrow \rangle$, correspond for fermions 
to determinants built out from the $\psi_j(k)\chi(\omega)$, $j=1,2,3$, spin orbitals, one finds that the 
general form of the many-body wave functions associated with the three vectors in Eq.\ (\ref{vv}) is
\begin{align}
\Psi_i = \sum_{l=1}^3 F_l^i(k_1,k_2,k_3) \zeta_l(\omega_1,\omega_2,\omega_3),
\label{psic}
\end{align} 
where the three spin primitives are given by $\zeta_1=\alpha(\omega_1)\alpha(\omega_2)\beta(\omega_3)$,
$\zeta_2=\alpha(\omega_1)\beta(\omega_2)\alpha(\omega_3)$, and 
$\zeta_3=\beta(\omega_1)\alpha(\omega_2)\alpha(\omega_3)$.

\begin{figure}[t]
\includegraphics[width=6.0cm]{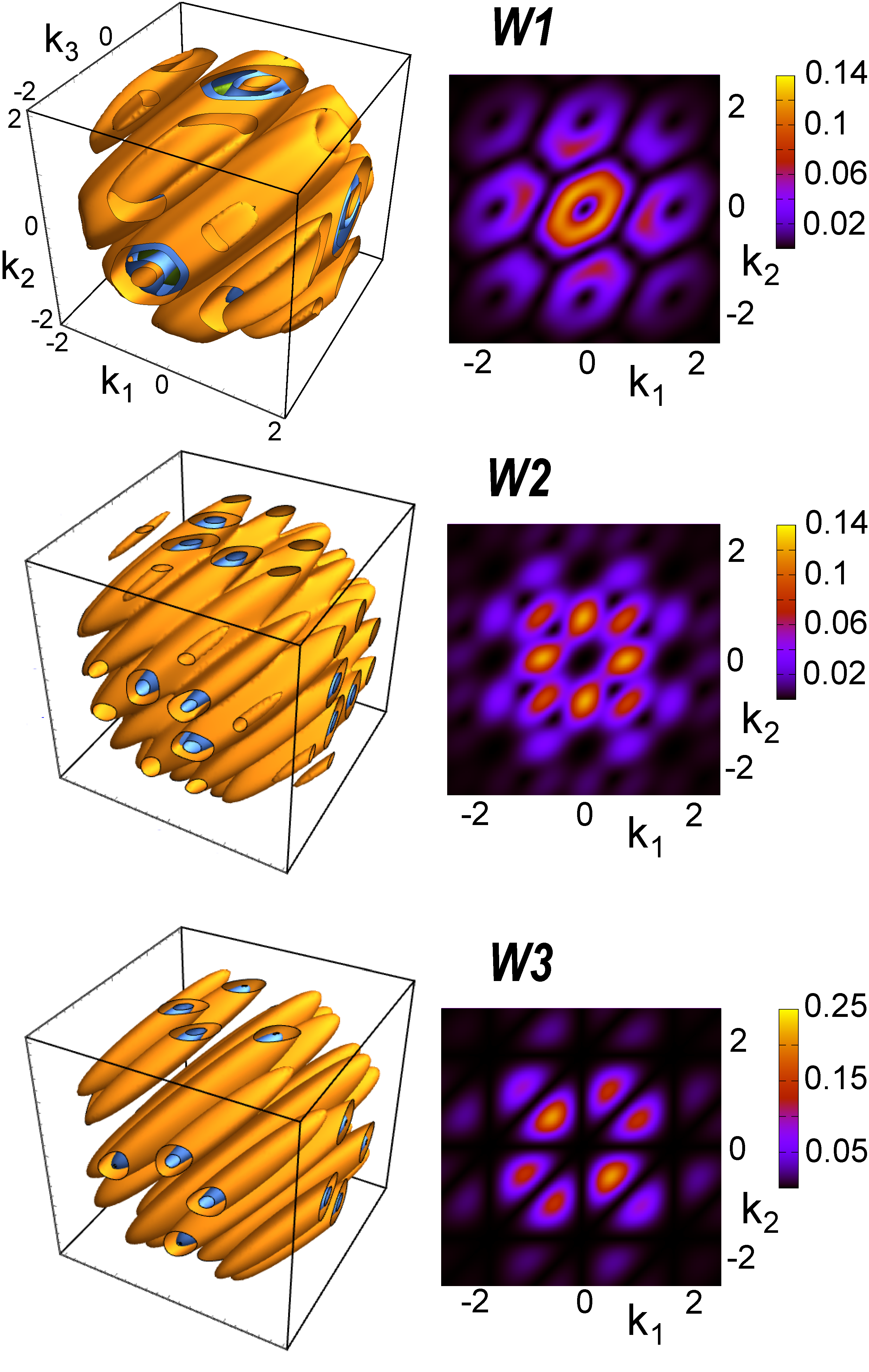}
\caption{Third-order spin-unresolved momentum correlations [see Eq.\ (\ref{g3un})] for the three $W$ states.
Left column: 3D contour plots. Right column: Corresponding 2D maps for cuts at $k_3=0$.
Top row: $W1$. Second row: $W2$. Third row: $W3$.
Parameters: $s=0.5$ $\mu$m and $D=3.8$ $\mu$m. Momenta in units in 1/$\mu$m. Third-order correlations in units
of $\mu$m$^3$.}
\label{fig1}
\end{figure}

\section{Results: The $W$ states}

\subsection{Spin {\it un\/}resolved third-order momentum correlations}

Since the spin primitive functions $\zeta_l$'s form an orthonormal set, one gets for the spin 
{\it unresolved\/} third-order correlations \cite{bran17} (i.e., summing over all possible spin cases using 
the formal integration over spins)
\begin{align}
\begin{split}
& {\cal G}_i^3(k_1,k_2,k_3) = \int \Psi_i^* \Psi_i d\omega_1 d\omega_2 d\omega_3 =\\
& \sum_{l=1}^3 |F_l^i(k_1,k_2,k_3)|^2.
\label{g3unf}
\end{split}
\end{align}

The calculations of the $F_l^i$'s out of the determinants are straighforward, but lengthy. We have used the
algebraic language MATHEMATICA \cite{math18} to carry them out. Below, we present the final analytic results. 

\begin{table}[b]
\caption{\label{tb1} Coefficients entering in Eq.\ (\ref{g3un}).}
\begin{ruledtabular}
\begin{tabular}{c|c|ccc}
$i$ & $E_i$   &  $A_i$  &  $B_i$  &  $C_i$  \\ \hline
3   &   0     & -2  & -1  & ~2  \\
2   & $-J/2$  & -1  & ~1  & -1  \\
1   & $-3J/2$ & ~1  & -1  & -1  \\
\end{tabular}
\end{ruledtabular}
\end{table} 

Assuming equal separations between the central and the outer wells (i.e., taking $d_1=-D$, $d_2=0$, $d_3=D$), 
the analytic expressions for the spin {\it unresolved\/} third-order momentum correlations corresponding to 
the three entangled $S_z=1/2$ Heisenberg states are given by the same general formula
\begin{align}
\begin{split}
& {\cal G}_i^3(k_1,k_2,k_3) = \frac{2\sqrt{2}}{3\pi^{3/2}} s^3 e^{-2(k_1^2+k_2^2+k_3^2)s^2} \{ 3 + \\
& \sum_{p < q}^3  A_i \cos[D(k_p-k_q)] + \sum_{p < q}^3  B_i \cos[2D(k_p-k_q)] + \\
& \sum_{(p,q,r)} C_i \cos[D(k_p+k_q-2k_r)] \},
\label{g3un} 
\end{split}
\end{align}   
where $(p,q,r)$ takes only the three values $(1,2,3)$, $(2,3,1)$, and $(3,1,2)$.
The associated coefficients $A_i$, $B_i$, and $C_i$ are given in TABLE \ref{tb1}.

Illustrations of the unresolved third-order momentum correlations for the three $W$ states in Eq.\ (\ref{vv})
are displayed in Fig.\ \ref{fig1}. The left column displays 3D isosurface contours, ${\cal G}_i^3(k_1,k_2,k_3)=
constant$, while the right column displays corresponding 2D cuts by keeping the third momentum fixed at $k_3=0$.
The plots illustrate visually that the three ${\cal G}_i^3(k_1,k_2,k_3)$ in Eq.\ (\ref{g3un}) exhibit 
sufficiently different map landscapes, which could be explored with experimental measurements.

Characteristic landscape patterns that allow differentiation between the $W$-states remain also prominent in 
the case of both spin-unresolved and spin-resolved second-order correlation maps, which are investigated next.

\subsection{Spin {\it un\/}resolved second-order momentum correlations}

When the $N$-particle many-body wave function $\Psi$ is available in the coordinate space, it is well-known 
that the $M$-order ($M \leq N)$ space correlations are obtained by carrying out the $N-M$ integrations 
of $\Psi^* \Psi$ over the remaining $M+1,M+2,\ldots,N$ variables \cite{lowd55,bran17}. 
In this case the corresponding $M$-order momentum correlations are determined via an appropriate Fourier 
transform of the space correlations \cite{bran17}. Here, the third-order correlations are already available in
momentum space at the very beginning; see Eqs.\ (\ref{g3unf}) and (\ref{g3un}). 
Thus the lower spin unresolved second-order correlations can be obtained simply from Eq.\ (\ref{g3un}) by 
integrating ${\cal G}_i^3$ over the third $k_3$ momentum variable. Then, neglecting the vanishing contributions
from the orbital overlaps (i.e., assuming $D^2/s^2 >> 1$), one finds:
\begin{align}
\begin{split}
& {\cal G}_i^2(k_1,k_2) = \int dk_3 {\cal G}_i^3(k_1,k_2,k_3) = \\
& \frac{2}{3\pi} s^2 e^{-2(k_1^2+k_2^2)s^2} \times \\
& \{ 3 +  A_i \cos[D(k_1-k_2)] +   B_i \cos[2D(k_1-k_2)] \},
\label{g2un}
\end{split}
\end{align}
where the coefficients $A_i$ and $B_i$ are the same as in TABLE I.

The spin-unresolved second-order correlations for the three $W$ states are plotted in the first column 
(for $W1$ and $W2$) and the fourth column, top row (for $W3$) of 
Fig.\ \ref{fig2}. It is characteristic that the main diagonal ($k_1-k_2=0$) acquires nonvanishing values for 
the two states with $S=1/2, S_z=1/2$ (i.e., for $W1$ and $W2$), while it exhibits vanishing values all along 
its extent for the third $(W3)$ state with $S=3/2,S_z=1/2$. Furthermore, the interference between the two 
length scales, $D$ and $2D$ [see Eq.\ (\ref{g2un})], generates a wavy doubling (cases of $W1$ and $W3$) or 
tripling (case of $W2$) of the dominant peaks of the fringes, which experimentally could be seen as broadening
of the fringes. Note that this wavy broadening of the fringes was reported in Ref.\ \cite{bran17} for the 
partial case of the $W1$ ground state.

\subsection{Spin resolved third-order correlations}

Spin resolved correlations impose specific values for the spins associated with the momenta variables $k_i$'s.
We note that knowledge of the spin resolved correlations provides a more complete degree of characterization 
of the many-body state compared to that obtained from knowledge of the spin unresolved correlations.

When the spins for all three momenta $k_i$'s are fixed, each vector solution in Eq.\ (\ref{vv}) allows 
three spin arrangements according to the three spin primitives $\zeta_1$, $\zeta_2$ and $\zeta_3$. As a result,
the following third-order three-spin resolved correlations for the three $Wi$, $i=1,2,3$, states can be 
specified:
\begin{align}
&{\cal G}_{\uparrow\uparrow\downarrow}^{3,i}(k_1,k_2,k_3)=|F_1^i(k_1,k_2,k_3)|^2,\\
&{\cal G}_{\uparrow\downarrow\uparrow}^{3,i}(k_1,k_2,k_3)=|F_2^i(k_1,k_2,k_3)|^2,\\
&{\cal G}_{\downarrow\uparrow\uparrow}^{3,i}(k_1,k_2,k_3)=|F_3^i(k_1,k_2,k_3)|^2.
\end{align}
   
The explicit analytic expressions (a total of nine) for the above three-spin resolved correlations, which are 
different from each other, are given in a compact form by the same general expression:

\begin{widetext}
\begin{align}
\begin{split}
&{\cal G}^{3,i}_{\rm spin-resolved}(k_1,k_2,k_3) = 
\frac{\sqrt{2}}{9 \pi^{3/2}} s^3 e^{-2(k_1^2+k_2^2+k_3^2)s^2} \times \\
& \big\{ 6  +  c_{12} \cos[D(k_1-k_2)] + c_{13} \cos[D(k_1-k_3)] + c_{23} \cos[D(k_2-k_3)]+ \\
& \widetilde{c}_{12} \cos[2D(k_1-k_2)] + \widetilde{c}_{13} \cos[2D(k_1-k_3)] + 
\widetilde{c}_{23} \cos[2D(k_2-k_3)]+ \\
& c_{123} \cos[D(k_1+k_2-2k_3)] +  c_{231} \cos[D(k_2+k_3-2k_1)] + 
c_{312} \cos[D(k_3+k_1-2k_2)] \big\},
\label{a1}
\end{split}
\end{align}
\end{widetext}
where the corresponding coefficients are listed in TABLE \ref{tbsm}.

\begin{table}[b]
\caption{\label{tbsm} Coefficients entering in the expression in Eq.\ (\ref{a1}) for the third-order 
spin resolved momentum correlations.}
\begin{ruledtabular}
\begin{tabular}{c|c|ccccccccc}
$W$-state & ${\rm spins}$ &  $c_{12}$  &  $c_{13}$  &  $c_{23}$ & 
$\widetilde{c}_{12}$ & $\widetilde{c}_{13}$ &  $\widetilde{c}_{23}$ &
$c_{123}$ & $c_{231}$ & $c_{312}$ \\ \hline
~~   & $\uparrow\uparrow\downarrow$ & -4 & -4 & -4 & -2 & -2 & -2 & ~4 & ~4 & ~4 \\
$W3$ & $\uparrow\downarrow\uparrow$ & -4 & -4 & -4 & -2 & -2 & -2 & ~4 & ~4 & ~4 \\ 
~~   & $\downarrow\uparrow\uparrow$ & -4 & -4 & -4 & -2 & -2 & -2 & ~4 & ~4 & ~4 \\\hline
~~   & $\uparrow\uparrow\downarrow$ & -6 & ~0 & ~0 & ~0 & ~3 & ~3 & -6 & ~0 & ~0 \\
$W2$ & $\uparrow\downarrow\uparrow$ & ~0 & -6 & ~0 & ~3 & ~0 & ~3 & ~0 & ~0 & -6 \\
~~   & $\downarrow\uparrow\uparrow$ & ~0 & ~0 & -6 & ~3 & ~3 & ~0 & ~0 & -6 & ~0 \\ \hline
~~   & $\uparrow\uparrow\downarrow$ & -2 & ~4 & ~4 & -4 & -1 & -1 & ~2 & -4 & -4 \\
$W1$ & $\uparrow\downarrow\uparrow$ & ~4 & -2 & ~4 & -1 & -4 & -1 & -4 & -4 & ~2 \\
~~   & $\downarrow\uparrow\uparrow$ & ~4 & ~4 & -2 & -1 & -1 & -4 & -4 & ~2 & -4 \\ 
\end{tabular}
\end{ruledtabular}
\end{table}

\subsection{Spin resolved second-order correlations}

We turn now to studying second-order spin resolved correlations. The \fulu spin resolved correlations for
the three $W$ states in Eq.\ (\ref{vv}) have the general form:
\begin{align}
\begin{split}
&{\cal G}_{\uparrow\uparrow}^{2,i}(k_1,k_2)=
\int dk_3 {\cal G}_{\uparrow\uparrow\downarrow}^{3,i}(k_1,k_2,k_3) =\\
& \frac{1}{9\pi} s^2  e^{-2(k_1^2+k_2^2)s^2} \times \\
& \{ 6 +  P_i \cos[D(k_1-k_2)] +   Q_i \cos[2D(k_1-k_2)] \},
\label{g2resuu}
\end{split}
\end{align}
where the coefficients $P_i$ and $Q_i$ are given in TABLE \ref{tb2}. Similarly, the other two
second-order spin resolved correlations, namely the \fuld, ${\cal G}_{\uparrow\downarrow}^{2,i}(k_1,k_2)=
\int dk_3 {\cal G}_{\uparrow\downarrow\uparrow}^{3,i}(k_1,k_2,k_3)$, and the \fdlu,
${\cal G}_{\downarrow\uparrow}^{2,i}(k_1,k_2)=
\int dk_3 {\cal G}_{\downarrow\uparrow\uparrow}^{3,i}(k_1,k_2,k_3)$ yield the same general form as in
Eq.\ (\ref{g2resuu}), with the specific values of the $P_i$ and $Q_i$ coefficients displayed in
TABLE \ref{tb2}.

\begin{table}[b]
\caption{\label{tb2} Coefficients for the second-order spin resolved momentum correlations
${\cal G}_{\uparrow\uparrow}^{2,i}(k_1,k_2)$, ${\cal G}_{\uparrow\downarrow}^{2,i}(k_1,k_2)$, 
and  ${\cal G}_{\downarrow\uparrow}^{2,i}(k_1,k_2)$ 
entering in Eq.\ (\ref{g2resuu}). The index $i$ counts the $W$ states in Eq.\ (\ref{vv}).}
\begin{ruledtabular}
\begin{tabular}{c|c|cc|cc|cc}
\multicolumn{2}{c|}{~~~} & \multicolumn{2}{c|}{$\uparrow\uparrow$} &
\multicolumn{2}{c|}{$\uparrow\downarrow$} & \multicolumn{2}{c}{$\downarrow\uparrow$} \\ \hline
$i$ & $E_i$   &  $P_i$  &  $Q_i$  &  $P_i$ & $Q_i$ & $P_i$ &  $Q_i$ \\ \hline
3   &   0     & -4  & -2 & -4  & -2  & -4  & -2  \\
2   & $-J/2$  & -6  & ~0 & ~0  & ~3  & ~0  & ~3  \\
1   & $-3J/2$ & -2  & -4 & ~4  & -1  & ~4  & -1  \\
\end{tabular}
\end{ruledtabular}
\end{table}

\begin{figure*}[t]
\includegraphics[width=12.0cm]{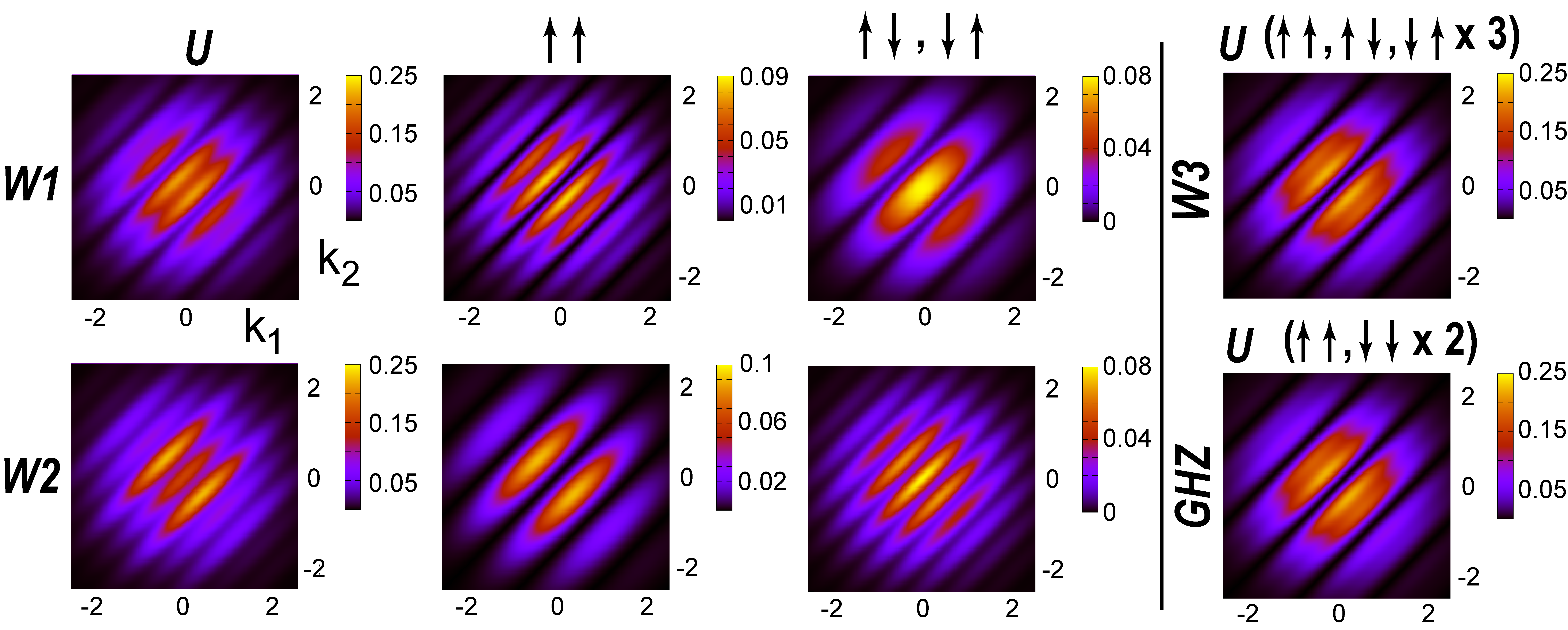}
\caption{Second-order momentum correlation maps for the $W$ and $GHZ$ states.
Top row: $W1$ state and $W3$ state (fourth column). Bottom row: $W2$ state and $GHZ$ state (fourth 
column). The spin-unresolved correlations are denoted by a $U$ (first and fourth column). Second and third 
column: Spin-resolved cases denoted by the symbols $\uparrow\uparrow$, $\uparrow\downarrow$, and 
$\downarrow\uparrow$. Other cases that coincide with the corresponding $U$ maps when multiplied by a factor of 
3 or 2 are indicated within the parentheses in the fourth column.
Parameters: $s=0.5$ $\mu$m and $D=3.8$ $\mu$m. Momenta in units in 1/$\mu$m. Second-order correlations in units
of $\mu$m$^2$.}
\label{fig2}
\end{figure*}

The second-order spin-resolved correlation maps for the two $W1$ and $W2$ states (with $S=1/2$) are displayed
in the second and third column of Fig.\ \ref{fig2}, respectively; for the $W3$ state (with $S=3/2$), see 
below. The \fuld and \fdlu maps for both states coincide, as indicated in the figure. The main 
diagonal in these maps ($k_1-k_2=0$) is associated with vanishing values (resulting in fringe valleys) for the
same-spin cases ($\uparrow \uparrow$), while it exhibits nonvanishing values (resulting in fringe ridges) 
for the different-spin cases ($\uparrow \downarrow$ or  $\downarrow \uparrow$); this is consistent with the
Pauli exclusion principle for same-spin fermions and the property that fermions with different spins are 
distinguishable. Furthermore, there is a clear contrast regarding the number of fringes for the spin-resolved
maps of the $W1$ and $W2$ states; indeed for the same-spin cases (second column of Fig.\ \ref{fig2}), there
are eight visisble fringes for $W1$ conmpared to only four visible fringes for $W2$. For the different-spin 
cases (third column of Fig.\ \ref{fig2}), the opposite trend appears, namely, there are only five visible
fringes for $W1$ compared to nine visible fringes for $W2$. Note that the sum of the three spin-resolved 
correlations equals the spin-unresolved one, symbolically $\uparrow\uparrow+\uparrow\downarrow+
\downarrow\uparrow = U$.

For the $W3$ case (with $S=3/2$, $S_z=1/2$), all three spin-resolved maps coincide. Each one of these maps
multiplied by a factor of three equals the spin-unresolved map; this is symbolically denoted at the top
of the frame situated on the top row, fourth column of Fig.\ \ref{fig2}. 

\section{Results: The $GHZ$ state} 

The $GHZ$ state is a linear superposition of the two fully polarized eigenstates of 
the Heisenberg Hamiltonian in Eq.\ (\ref{hh}), that is, 
\begin{align}
|GHZ\rangle = (|\uparrow \uparrow \uparrow\rangle + |\downarrow \downarrow \downarrow \rangle)/\sqrt{2}.
\label{ghz}
\end{align}

The corresponding energy is $E_{\rm GHZ}=0$ and the total spins are $S=3/2$ 
(a spin eigenvalue) and $\langle S_z \rangle=0$ (an expectation value, not a spin eigenvalue). 
The second-order spin-unresolved correlation map for the $GHZ$ state is displayed in Fig.\ \ref{fig2}
(second row, fourth column).
It is immediately seen that the $GHZ$ spin-unresolved map coincides with that of
the $W3$ spin-unresolved map displayed also in Fig.\ \ref{fig2}, top of fourth column. This
result was also explicitly verified by deriving via our methodology the 
corresponding analytic $GHZ$ expression and comparing it with that in Eq.\ (\ref{g2un}) (for $i=3$). Namely 
starting from the associated determinants for the two $|\uparrow \uparrow \uparrow\rangle$ and 
$|\downarrow \downarrow \downarrow \rangle$ kets in Eq.\ (\ref{ghz}), we calculated first the third-order 
$GHZ$ momentum correlations and subsequently we derived the second-order correlations through an integration 
over the third momentum $k_3$ variable. Furthermore, the $GHZ$ second-order spin-resolved correlation maps,
$\uparrow \uparrow$ and $\downarrow \downarrow$, coincide and equal the spin-unresolved one when multiplied
by a factor of two. Finally and consistent with the above, we found through our analytic calculations
(not shown) that the $GHZ$ third-order spin-unresolved correlation maps coincide with those associated 
separately with each fully polarized state $|\uparrow \uparrow \uparrow\rangle$ ($S=3/2$, $S_z=3/2$) or 
$|\downarrow \downarrow \downarrow \rangle$ ($S=3/2$, $S_z=-3/2$), as well as with that of the $W3$ state
(which also has $S=3/2$); see Eq.\ (\ref{g3un}), for $i=3$. 
 
\section{Conclusions}

Analytical expressions for the third-order and second-order spin-resolved and 
spin-unresolved momentum correlations for the strongly-entangled $W$ and $GHZ$ states \cite{zeil90,cira00}
of three singly-trapped ultracold fermionic atoms have been derived. The associated correlation patterns and 
maps are related \cite{yann19} to nowadays experimentally accessible TOF measurements; they enable matter-wave 
interference studies in analogy with recent three-photon interferometry \cite{agne17,mens17,tamm15,tamm18}. 
A main finding is that knowledge of the spin-unresolved correlation maps is required to fully characterize the 
strongly-entangled states. 

\textcolor{black}{
This work uncovers and demonstrates a methodology which allows treatment of strongly interacting entangled states  
which are outside the scope of the standard Wick's factorization scheme \cite{hodg11,aspe19,prei19}, thus opening 
the door and providing the impetus for experimental investigations, using coincidence time-of-flight measurements 
on trapped ultracold atom systems, of entangled states (like the $W$ and $GHZ$ ones treated here) which are 
ubiquitous in quantum information theory and protocols in quantum communication and cryptography and studies of the
fundamentals of quantum mechanics \cite{gree89}.
}

\section{Acknowledgments} 
This work has been supported by a grant from the Air Force Office of Scientic Research 
(AFOSR, USA) under Award No. FA9550-15-1-0519. Calculations were carried out at the GATECH Center for 
Computational Materials Science.


\begin{thebibliography}{99}
\bibitem{note2}
After submission of our paper, an experimental study appeared (see Ref.\ \cite{berg19}), where the results of 
time-of-flight measurements of second-order position and momentum correlations for entangled 
Einstein-Podolsky-Rosen states of two ultracold fermionic atoms have been presented.
\bibitem{note1}
After submission of our paper, a second experimental study has also appeared (see Ref.\ \cite{prei19}), 
where the Wick's theorem is utilized for analysis of time-of-flight measurements of a non-interacting 
spin-polarized state (unentangled, single determinant, highest spin-component) of optical-tweezer-trapped three 
$^6$Li ultracold atoms.
\end{thebibliography}
\end{document}